\documentclass[10pt,conference]{IEEEtran}
\usepackage{siunitx}
\usepackage{graphicx}
\usepackage{tabularx}
\usepackage{float}
\usepackage{algorithm}
\usepackage{bm}
\usepackage{algpseudocode}
\usepackage{amsfonts}
\usepackage{amsmath}
\usepackage{indentfirst}
\usepackage{mathtools}
\usepackage{listings}
\usepackage{courier}
\usepackage{amssymb}
\usepackage{xcolor}

\begin{document}
\title{Interference Alignment with Power Splitting Relays in Multi-User Multi-Relay Networks}

\author{\IEEEauthorblockN{Man Chu\IEEEauthorrefmark{1}, Biao He\IEEEauthorrefmark{2}, Xuewen Liao\IEEEauthorrefmark{1}, Zhenzhen Gao\IEEEauthorrefmark{1}, and Shihua Zhu\IEEEauthorrefmark{1}}
\IEEEauthorblockA{\IEEEauthorrefmark{1}Department of Information and Communication Engineering, Xi'an Jiaotong University, China\\
\IEEEauthorrefmark{2}
The Center for Pervasive Communications and Computing, University of California at Irvine, Irvine, CA 92697, USA}
Email: cmcc\_1989414@stu.xjtu.edu.cn, biao.he@uci.edu, \{yeplos, zhenzhen.gao, szhu\}@mail.xjtu.edu.cn\vspace{5mm}
}
\maketitle

\begin{abstract}
In this paper, we study a multi-user multi-relay interference-channel network, where energy-constrained relays harvest energy
from  sources' radio frequency (RF) signals and use the harvested energy to forward the information to destinations.
We adopt the interference alignment (IA) technique to address the issue of interference, and propose a novel transmission scheme with the IA at sources  and  the power splitting (PS) at relays. A distributed and iterative algorithm to obtain the optimal PS ratios is further proposed, aiming at maximizing the sum rate of the network. The analysis is then validated by simulation results. Our results show that the proposed scheme with the optimal design significantly improves the performance of the network.
%
%
\end{abstract}

\vspace{2mm}
\section{Introduction}
Energy harvesting has been envisioned as a  promising technique to provide perpetual power supplies and prolong the lifetime of energy-constrained wireless networks, since wireline charging and battery replacement are not always  feasible or unhazardous.
In particular, simultaneous wireless information and power transfer (SWIPT) has drawn a significant amount of attention with its advantage of transporting both energy and information at the same time through radio frequency signals.



The concept of SWIPT was first put forward by Varshney in his pioneer work~\cite{1Varshney}, in which he investigated  the tradeoff between information delivery and energy transfer.
Recently developed practical receiver designs for SWIPT include power splitting (PS) and time switching (TS)~\cite{2RuiZhang}.
For TS, the receiver switches over time between information decoding and energy harvesting. Differently, the receiver with PS splits the received signal power between information decoding and energy harvesting.
It has been shown that the TS is theoretically a special case of the PS
with binary PS ratios. Thus, the PS usually outperforms the TS in terms of the  rate-energy tradeoffs.
With the in-depth study on SWIPT, the research problem of how to combine SWIPT with other key technologies has gained much attention~\cite{3Ng}. For the combination of the SWIPT and relays, both single-relay and multi-relay networks have been considered to acquire the efficient SWIPT scheme in~\cite{4ChenZ}\cite{5HuL}.
In addition, the interference channel (IC) has been taken into account for the
SWIPT with relays in, e.g.,~\cite{6Krikidis} and~\cite{7ChenH}.
The authors in~\cite{6Krikidis} analyzed the outage performance for SWIPT both with and without relay coordination in multi-user IC networks. In~\cite{7ChenH}, the SWIPT with relays in IC networks was studied from a game-theoretic perspective.
However, none of the aforementioned studies has directly addressed the issue of  interference for SWIPT with relays in IC networks. In practice, interference alignment (IA) is recognized as an efficient technique to address the issue of interference for wireless transmissions. With the IA, the interferences are aligned into certain subspaces of the effective channel to the receiver, and the desired signals are aligned into interference-free subspaces~\cite{8HZeng}. 
Without the consideration of relays, the authors in~\cite{9LiX} studied the antenna selection with IA for the SWIPT in the IC network. While to the best of authors' knowledge, few work has been investigated in the literature on SWIPT with the IA in multiple AF (amplify and forward) relay networks.

In this paper, we study an IC network where multiple sources  transmit messages to multiple users through multiple AF relays, where the energy-constrained relays harvest energy from the RF signals and use that harvested energy to forward the information to destinations.
We propose a novel transmission scheme by adopting the IA and the PS design.
We further propose a distributed and iterative algorithm to obtain the optimal PS ratios at relays that maximize the sum rate of the network.
Our results show that the proposed scheme effectively improves the performance
of the network, and it significantly outperforms benchmark schemes.

\vspace{2mm}
\section{System Model}
\begin{figure}[!t]
\centering
\includegraphics[width=0.7\columnwidth]{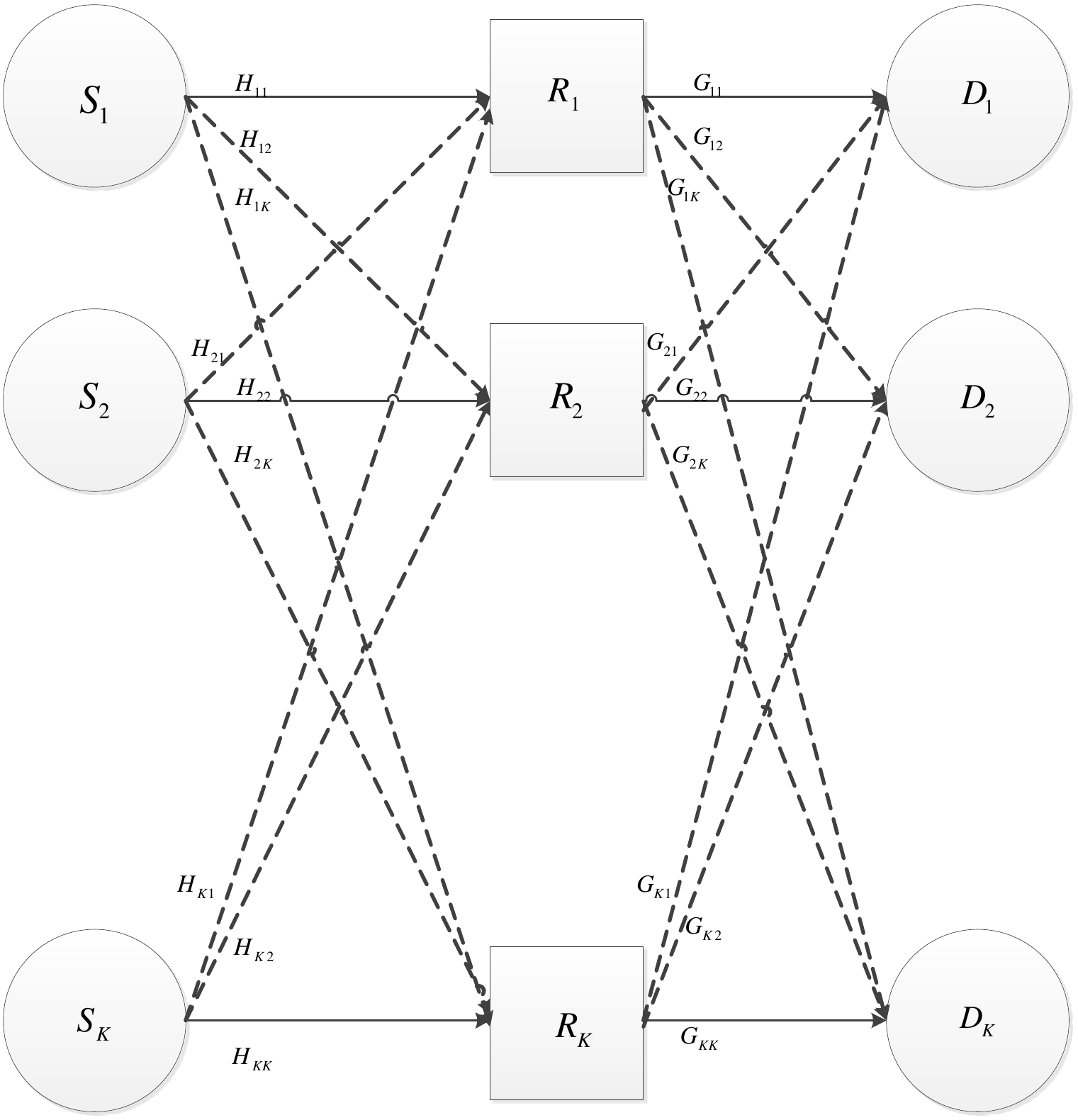}
\caption{Illustration of multi-user multi-relay interference MIMO network.}
\end{figure}
We consider a $K\times K\times K$ dual-hop IC network consisting of $K$ sources, $K$ relays, and $K$ destinations, as shown in Figure~1. Each source $S_i$ intends to transmit messages to its respective destination $D_i$ with the help of its dedicated AF relay $R_i$, $i\in\left\{1, \cdots, K\right\}$. We assume that there are no direct links between the sources and the destinations.
The numbers of antennas at each source, relay, and destination are denoted by $M$, $N$, and $L$, respectively.
The sets of sources, relays and destinations are denoted by $\mathcal{N}_S$, $\mathcal{N}_R$, and $\mathcal{N}_D$, respectively.
In addition, the set of all source-relay-destination ($S$-$R$-$D$) links, $S_i\rightarrow R_i \rightarrow D_i$, $i=1, \cdots K$, is denoted by~$\mathcal{K}$.

We adopt the quasi-static block Rayleigh fading model and assume that all channels are independent of each other.
The $N\times M$ normalized channel matrix from $S_i$ to $R_j$ is denoted by $\mathbf{H}_{ij}$, and the $L\times N$ normalized channel matrix from $R_i$ to $D_j$ is denoted by $\mathbf{G}_{ij}$, $i,j\in\left\{1, \cdots, K\right\}$.
We assume that each node has the local channel state information (CSI) to enable the distributed IA.
Moreover, the distance between $S_i$ and $R_j$ is denoted by $r_{ij}$, and the distance between $R_i$ and $D_j$ is denoted by $m_{ij}$. The correspondingly path-loss effects are modelled by $r_{ij}^{-\tau}$ and $m_{ij}^{-\tau}$, respectively, where $\tau\ge2$ denotes the path-loss exponent~\cite{10Nasir_A}.

Half-duplex relays are considered in this work, and the two hop transmissions operate in two time slots. In the first slot, the sources simultaneously transmit the information and energy to the relays. In the second slot, the relays forward the received information to the destinations by the harvested power in the first slot. We assume that the wireless power transfer is the only energy access for the relays. Besides, every time slot is assumed to be unit interval for simplicity.


\vspace{2mm}
\section{Interference Alignment with Power-Splitting Relays}
In this section, we proposed the transmission scheme for the dual-hop IC with the IA and the PS relays.

\subsection{IA Between Sources and Relays}
In the first time slot, the IA is employed for the transmissions from sources to relays. That is, all interference signals from unintended sources are aligned into certain interference subspaces at any relay $R_i$.

As per the mechanism of IA,  the following conditions should be satisfied to have the perfect IA:\vspace{2mm}
\begin{align}
{{\mathbf{U}_i^H}}{\mathbf{H}_{ji}}{\mathbf{V}_j}={\bm{0}},{\forall{j\neq{i}}},
\end{align}
\begin{align}
\mathrm{rank}({{\mathbf{U}_i^H}}{\mathbf{H}_{ii}}{\mathbf{V}_i})={d_i}
\end{align}
\begin{align}
{M+N}\geq(K+1)\times{d_i},
\end{align}
where ${\mathbf{V}_i}\in{\mathbb{C}}^{{M}\times{d_i}}$ denotes the unitary precoding matrix at $S_i$,  ${\mathbf{U}_i}\in{\mathbb{C}}^{{N}\times{d_i}}$ denotes the unitary decoding matrix at $R_i$, and $d_i$ denotes the number of data streams from $S_i$.
Here, the perfect IA means that the interferences are aligned into the null space of the effective channel to the relay and can be eliminated completely.
We can interpret (1) as the condition of having the interference-free space of the desired dimensions, and (2) and (3) as the conditions that the desired signal is visible and resolvable within the interference-free space~\cite{11Gomadam}.
When the number of links is larger than 3, i.e., $K>3$, the closed-form expressions for the precoding and decoding matrixes cannot be obtained. Hence, we use the channel reciprocity characteristic with the distributed and iterative IA to determine the IA matrices in this work.

The received signal at $R_i$ (without  multiplexing the decoding matrix ${\mathbf{U}_i^H}$) in the first time slot is given by
\begin{align}
\bm{y}_{R_i}^{IA}=&\sqrt{p_i r_{ii}^{-\tau}}{\mathbf{H}_{ii}}{\mathbf{V}_i}{\bm{x}_i}\nonumber\\
&+{\sum_{j=1,j\neq{i}}^K\sqrt{p_jr_{ji}^{-\tau}}}{\mathbf{H}_{ji}}{\mathbf{V}_j}{\bm{x}_j}+{\bm{n}_{A_i}},
\end{align}
where $p_i$ denotes the transmit power at $S_i$, ${\bm{x_i}}\in{\mathbb{C}}^{{d_i}\times{1}}$ denotes the normalized desired signal for $D_i$ with $\mathbb{E}\{||\bm{x}_i||^2\}=1$, and $\bm{n}_{A_i}$ denotes the additive white Gaussian noise (AWGN) at the receiver antennas of $R_i$.
\newcounter{MYtempeqncnt}
\begin{figure*}[!t]
\normalsize
\setcounter{MYtempeqncnt}{\value{equation}}
\setcounter{equation}{11}
\begin{equation}
\label{eqn_dbl_x}
\bm{\gamma}_{D_i}{\left(\bm{\rho}\right)}=\frac{{\left({1-\rho_i}\right)}{X_i}{\rho_i}{p_i}{Y_{ii}}{\frac{1}{{\rho_i}p_i{b_{ii}}+{\sigma^2}}}}
{\sum_{j=1,j\neq{i}}^{K}{\left({1-\rho_j}\right)}{X_j}{\frac{{\rho_j}{p_j}{Y_{ji}}+c_{ji}{\sigma^2}}{\rho_j{p_j}{b_{jj}}+\sigma^2}}
+\left({1-\rho_i}\right){X_i}{c_{ii}}{\frac{\sigma^2}{\rho_ip_i{b_{ii}}+\sigma^2}}+\frac{\sigma^2}{\eta}}.
\end{equation}
\setcounter{equation}{\value{MYtempeqncnt}}
\hrulefill
\vspace*{4pt}
\end{figure*}

\subsection{PS Relays}
\begin{figure}[!t]
\centering
\includegraphics[width=0.99\columnwidth]{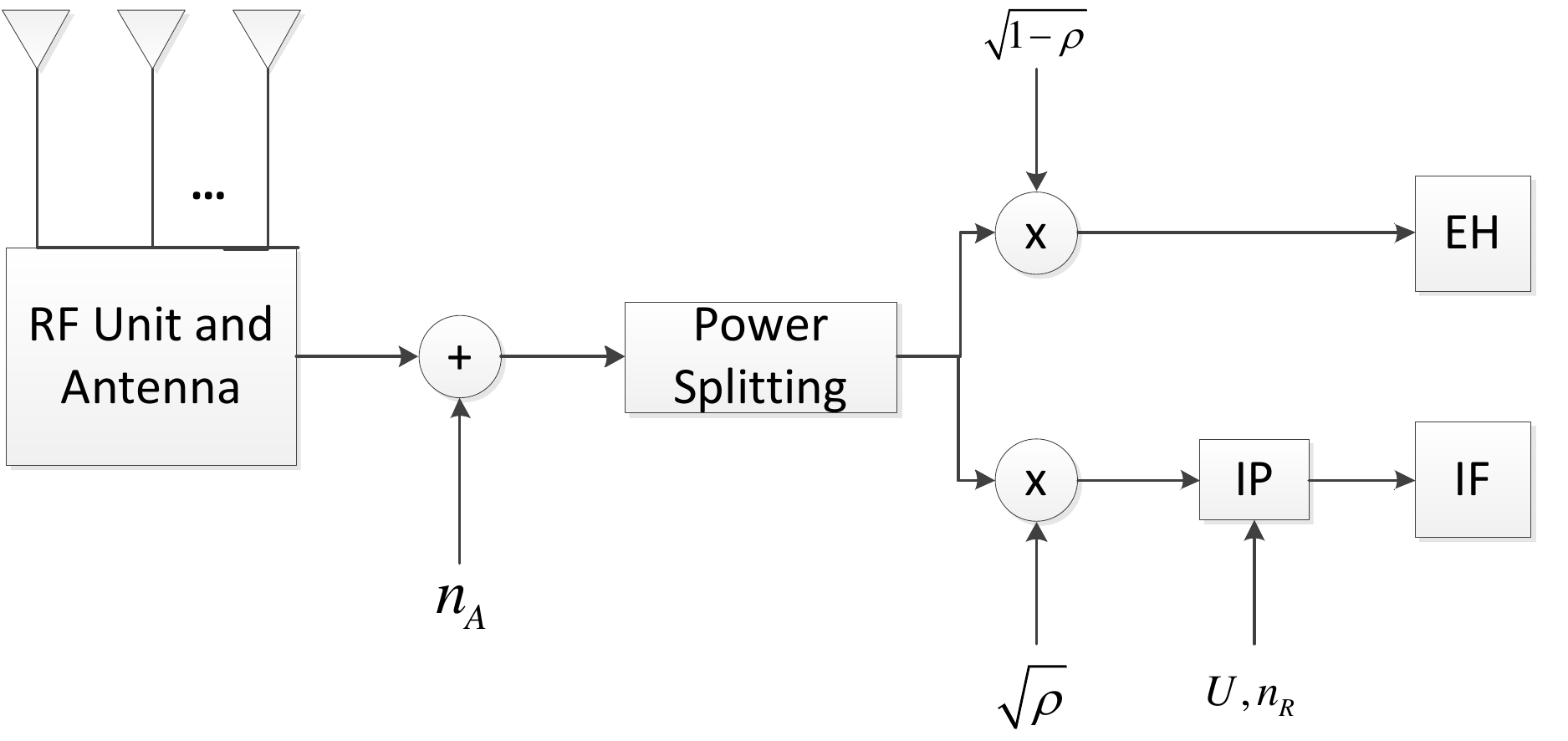}
\caption{Power splitting model at the relay.}
\label{Fig:psmode}
\end{figure}
In this work, the PS scheme is adopted at the relays for the SWIPT.
The block diagram of the receiver at the relay is shown in Figure~2. 
For the received signal at $R_i$, a fraction $0\le \rho_i\le1$ of the power is used for information processing and forwarding, and the remaining fraction $1-\rho_i$ of the power is used for energy harvesting. We refer to $\rho_i$ as the PS ratio at $R_i$.
We assume that the processing power required by the transmit/receive circuits at the relay is negligible compared with the power used for signal transmission~\cite{10Nasir_A}.

Then, the signals received by the energy-harvesting unit at $R_i$ is given by
\begin{eqnarray}
\bm{y}_{R_i}^{EH}&=&{\sqrt{(1-\rho_{i})}}\left(\sqrt{p_ir_{ii}^{-\tau}}\mathbf{H}_{ii}{\mathbf{V}_i}{\bm{x}_i}\right.\nonumber\\
&&+{\sum_{j=1,j\neq{i}}^K}\left.\sqrt{p_jr_{ji}^{-\tau}}\mathbf{H}_{ji}\mathbf{V}_{j}\bm{x}_{j}+{\bm{n}}_{A_i}\right),
\end{eqnarray}
and the harvested energy at $R_i$ is given by
\begin{equation}
Q_{R_i}={\eta}\left(1-{\rho_i}\right)\left({\sum_{j=1}^K}{p_j}{r_{ji}^{-\tau}}{||{\mathbf{H}_{ji}}{\mathbf{V}_j}||}^2\right),
\end{equation}
where $\eta$ denotes the energy conversion efficiency. Note that the harvested energy due to the noise $\bm{n}_{A_i}$ is very small and thus ignored.
\newcounter{MYtempeqncnt2}
\begin{figure*}[!t]
\normalsize
\setcounter{MYtempeqncnt2}{\value{equation}}
\setcounter{equation}{15}
\begin{equation}
\frac{\partial{\bm\gamma_{D_i}(\bar{\bm{\rho_i}},\rho_i)}}{\partial{\rho_i}}=
\frac{\left(z_2z_3+t\right){\rho_i}^3-\left(z_3p_i{b_{ii}}+z_2z_3{\sigma}^2-2t\right){\rho_i}^2+\left(2z_1z_3-z_2z_3
{\sigma}^2+t+p_i{b_{ii}}{\sigma}^2\right){\rho}^2+z_1z_3}
{\left[z_2+{\eta}\left(1-\rho_i\right)X_i{c_{ii}}{\sigma}^2\left({\rho_i}p_i{b_{ii}}\right)^{-1}+{\sigma}^2\right]^2}=0.
\vspace{3mm}
\end{equation}
\hrulefill
\setcounter{equation}{\value{MYtempeqncnt2}}
\end{figure*}
\newcounter{MYtempeqncnt3}
\begin{figure*}[!t]
\normalsize
\setcounter{MYtempeqncnt3}{\value{equation}}
\setcounter{equation}{16}
\begin{equation}
z_1=\sum\limits_{j=1,j\neq{i}}^K{\eta}\left(1-\rho_j\right)X_j\left({p_i}{b_{ii}}
+\sigma^2\right)\frac{\rho_j{p_j}Y_{ji}+{c_{ji}}\sigma^2}{\rho_j{p_j}{b_{ji}}+\sigma^2}-X_i{c_{ii}}\sigma^2+p_i{b_{ii}}\sigma^2.
\vspace{3mm}
\end{equation}
\begin{equation}
z_2=\sum\limits_{j=1,j\neq{i}}^K{\eta}\left(1-\rho_j\right)X_jY_{ji}{\frac{\rho_j{p_j}}{\rho_j{p_j}{b_{jj}}+\sigma^2}}+
\sum\limits_{j=1,j\neq{i}}^K{\eta}\left(1-\rho_j\right)X_j{c_{ji}}{\frac{\sigma^2}{\rho_j{p_j}{b_{jj}}+\sigma^2}}.
\end{equation}
\setcounter{equation}{\value{MYtempeqncnt3}}
\hrulefill
\vspace*{4pt}
\end{figure*}

At the information processing unit of $R_i$, the decoding matrix ${\mathbf{U}_i^H}$ is multiplexed to the received signal to eliminate the interferences.  The processed signals at the information processing unit of $R_i$ is then given by
\begin{equation}
\bm{y}_{R_i}^{IP1}=\sqrt{\rho_ip_ir_{ii}^{-\tau}}{\mathbf{U}_i^H}{\mathbf{H}_{ii}}{\mathbf{V}_i}{\bm{x}_i}+
{\sqrt{\rho_i}}{\mathbf{U}_i^H}{\bm{n}_{A_i}}+\bm{n}_{R_i},
\end{equation}
where $\bm{n}_{R_i}\sim\mathcal{CN}(\mathbf{0}, \sigma^2_{R_i}\mathbf{I}_N)$ denotes the AWGN introduced by the information processing unit at the relay. In practice, the power of antenna noise $\bm{n}_{A_i}$ is usually much smaller than the processing noise, and hence, the antenna noise has negligible impact on the signal processing. Then, the antenna noise $\bm{n}_{A_i}$ is ignored for simplicity in the rest of analysis in the paper~\cite{7ChenH,12LiuWIPTA}, and the processed signal at $R_i$ is simplified as
\begin{equation}
\bm{y}_{R_i}^{IP2}=\sqrt{\rho_ip_ir_{ii}^{-\tau}}{\mathbf{U}_i^H}{\mathbf{H}_{ii}}{\mathbf{V}_i}{\bm{x}_i}+
\bm{n}_{R_i}.
\end{equation}

\subsection{AF Signals to Destinations}
In the second time slot, the relays amplify and forward the processed signals to destinations by the harvested energy in the first time slot.
Since the utilization of IA for the second hop would significantly increase the system overhead and the energy consumption at the energy-constrained relays in practice, the IA is not adopted for the second-hop transmission.

The transmitted signal at $R_i$ in the second time slot is then given by
\begin{eqnarray}
\bm{y}_{R_i}^{F}=\sqrt{Q_{R_i}}{\beta_{R_i}}{\bm{y}_{R_i}^{IP2}},
\end{eqnarray}
where $Q_{R_i}$ denotes the transmit power of $R_i$ as given in (6) and
 \begin{eqnarray}
\beta_{R_i}=\frac{1}{\sqrt{{\rho_i}{p_i}{r_{ii}^{-\tau}}{||{\mathbf{U}_i^H}\mathbf{H}_{ii}\mathbf{V}_i||}^2+{\sigma_{R_i}^2}}}
\end{eqnarray}
denotes the power normalization coefficient at  $R_i$.
The received signal at $D_i$ is given by
\begin{eqnarray}
\bm{y}_{D_i}&=&{\sqrt{Q_{R_i}{m_{ii}^{-\tau}}}}{\mathbf{G}_{ii}}{\beta_{R_i}}{\bm{y}_{R_i}^{IP2}}\nonumber \\
&&+\sum_{j=1,j\neq{i}}^K{\sqrt{Q_{R_j}{m_{ji}^{-\tau}}}}{\mathbf{G}_{ji}}{\beta_{R_j}}{\bm{y}_{R_j}^{IP2}}\nonumber+\bm{n}_{D_i}\nonumber\\
&=&{\sqrt{Q_{R_i}{m_{ii}^{-\tau}}}}{\mathbf{G}_{ii}}{\beta_{R_i}}{}{\sqrt{\rho_ip_i{r_{ii}^{-\tau}}}}{\mathbf{U}_i^H}{\mathbf{H}_{ii}}{\mathbf{V}_i}{\bm{x}_i}\nonumber\\
&&+\sum_{j=1,j\neq{i}}^K{\sqrt{Q_{R_j}{m_{ji}^{-\tau}}}}{\mathbf{G}_{ji}}{\beta_{R_j}}{}{\sqrt{\rho_jp_j{r_{jj}^{-\tau}}}}{\mathbf{U}_j^H}{\mathbf{H}_{jj}}\nonumber\\
&&{\mathbf{V}_j}{\bm{x}_j}+\sum_{j=1}^K{\sqrt{Q_{R_j}{m_{ji}^{-\tau}}}}{\mathbf{G}_{ji}}{\beta_{R_j}}{\bm{n}_{R_j}}+\bm{n}_{D_i},
\end{eqnarray}
where $\bm{n}_{D_i}\sim\mathcal{CN}(\mathbf{0}, \sigma^2_{D_i}\mathbf{I}_N)$ denotes the AWGN at $D_i$.

In order to simplify the expression, we define $X_i=\sum_{j=1}^K{p_j}{r_{ji}^{-\tau}}{||{\mathbf{H}_{ji}\mathbf{V}_j}||}^2$, $Y_{ii}={m_{ii}^{-\tau}}{r_{ii}^{-\tau}}{||{\mathbf{G}_{ii}}{\tilde{\mathbf{H}}_{ii}}||}^2$, $Y_{ji}={m_{ji}^{-\tau}}{r_{ii}^{-\tau}}{||{\mathbf{G}_{ji}}{\tilde{\mathbf{H}}_{jj}}||}^2$, $b_{ii}={r_{ii}^{-\tau}}{||{\tilde{\mathbf{H}}_{ii}}||}^2$, $c_{ji}={m_{ji}^{-\tau}}{||\mathbf{G}_{ji}||}^2$, and the effective channel matrix $\tilde{\mathbf{H}}_{ii}={\mathbf{U}_i^H}{\mathbf{H}_{ii}}\mathbf{V}_i$.
In addition, we assume that $\sigma_{D_i}^2=\sigma_{R_i}^2=\sigma^2$ for any $i\in\{1,\cdots,K\}$, without loss of generality~\cite{7ChenH}.
The signal to interference plus noise power ratio (SINR) at $D_i$, denoted by $\bm\gamma_{D_i}(\bm{\rho})$ is then given by (12) as shown at the top of this page, where $\bm{\rho}=\left[\rho_1,\rho_2,\cdots,\rho_K\right]^{T}$ denotes the vector of all relays' PS ratios.
Finally, the rate of the transmission to $D_i$ is given by $\frac{1}{2} B \log_2{(1+\bm\gamma_{D_i}(\bm{\rho})})$, where $B$ denotes the frequency bandwidth and $1/2$ is due to the use of two time slots.

\vspace{2mm}
\section{Optimization of Power Splitting Ratio}
We note that the PS ratio $\bm{\rho}$ plays a pivotal in determining the network performance. Thus, we optimize the design of $\bm{\rho}$ in this section.

\subsection{Problem Formulation}
The objective of our design here is to determine a series of PS ratios, $\rho_1, \cdots, \rho_K$, to maximize the total transmission rate of the network.
The sum-rate maximization problem is formulated as:
\begin{align}
\left(\text{P1}\right)~&\max\limits_{\bm{\rho}}\;{R_{\mathrm{sum}}}(\bm{\rho})= \sum\limits_{i = 1}^K{\frac{1}{2}B\log_2\left(1+\bm\gamma_{D_i}(\bm{\rho})\right)},\nonumber \\
&\;\;{\rm{s}}{\rm{.t}}{\rm{.}}\;\;\;\;{\bm{\rho}}\in{\mathcal{A}},\nonumber\\
&\;\;\;\;\;\;\;\;\;\;{\mathbf{U}_i}{\mathbf{U}_i^H}=\mathbf{I}, {\mathbf{V}_i}{\mathbf{V}_i^H}=\mathbf{I},\nonumber\\
&\;\;\;\;\;\;\;\;\;\;\mathbb{E}\{||\bm{x}_i||^2\}=1, \forall{i\in\{1,\cdots,K\}},
\setcounter{equation}{12}
\end{align}
where $\mathcal{A}=\left\{\bm\rho\mid{0\leq{\rho_i}\leq{1}},\forall{i\in\{1,\cdots,K}\}\right\}$ is the feasible region of all PS ratios.

We find that the optimization problem in (13) is not convex, and the closed-form solution of the globally optimal $\bm{\rho}$ cannot be obtained.
Thus, we develop  a distributed and iterative scheme, such that each $R_i$
decides its own PS strategies aiming at maximizing the individual achievable transmission rate through the $i^{\text{th}}$ link, i.e., $S_i\rightarrow R_i \rightarrow D_i$, and the global optimal $\bm{\rho}$ that maximizes the sum rate of all links can be then obtained iteratively by the cooperation between the relays.
The design problem is formulated as:
\begin{align}
\left(\text{P2}\right)~&\sum\limits_{i=1}^K\;\max \limits_{\bm{\rho}}\;{\frac{1}{2}B\log_2\left(1+\bm\gamma_{D_i}(\bar{\bm{\rho_i}},\rho_i)\right)},\nonumber \\
&\;\;\;\;\;\;\;\;{\rm{s}}{\rm{.t}}{\rm{.}}\;{\bm{\rho}}\in{\mathcal{A}},\nonumber\\
&\;\;\;\;\;\;\;\;\;\;\;\;\;{\mathbf{U}_i}{\mathbf{U}_i^H}=\mathbf{I}, {\mathbf{V}_i}{\mathbf{V}_i^H}=\mathbf{I},\nonumber\\
&\;\;\;\;\;\;\;\;\;\;\;\;\;\mathbb{E}\{||\bm{x}_i||^2\}=1, \forall{i\in{\{1,\cdots,K}\}},
\end{align}
where $\bar{\bm{\rho_i}}=\{\rho_j\mid{1\leq{j}\leq{K}, \forall{j\neq{i}}}\}$ denotes the vector of all links' PS ratios except for the $i^{\text{th}}$ link.

\subsection{Optimal Design}
We first determine the optimal $\rho_i$ at $R_i$ that maximizes the received SINR at $D_i$ for any given $\bar{\bm{\rho_i}}$. The problem is formulated as:
\begin{align}
&\max\limits_{{\rho_i}}~\bm\gamma_{D_i}(\bar{\bm{\rho_i}},\rho_i),\nonumber \\
&\;\;{\rm{s}}{\rm{.t}}{\rm{.}}\;\;{\bm{\rho}}\in{\mathcal{A}},\nonumber\\
&\;\;\;\;\;\;\;\;{\mathbf{U}_i}{\mathbf{U}_i^H}=\mathbf{I}, {\mathbf{V}_i}{\mathbf{V}_i^H}=\mathbf{I},\nonumber\\
&\;\;\;\;\;\;\;\;\mathbb{E}\{||\bm{x}_i||^2\}=1, \forall{i\in\{1,\cdots,K\}}.
\end{align}
We find that $\bm\gamma_{D_i}|_{\rho_i=0}=0$ and $\bm\gamma_{D_i}|_{\rho_i=1}=0$, and hence, $\gamma_{D_i}(\bar{\bm{\rho_i}},\rho_i)$ is not a monotonous function of $\rho_i$ in the range of $[0,1]$.
We will later show the uniqueness of the extreme value point of $\gamma_{D_i}(\bar{\bm{\rho_i}},\rho_i)$ for $0\le\rho_i\le1$, and verify that the extreme point is the optimal PS ratio maximizing the SINR.


Let the first-order derivative of $\bm\gamma_{D_i}(\bar{\bm{\rho_i}},\rho_i)$ with respect to $\rho_i$ be equal to zero, we have (16) shown at the top of this paper, where $t=X_i{c_{ii}}{\sigma}^2$, $z_3=X_ip_iY_{ii}$, and $z_1$
and $z_2$ are given in (17) and (18), respectively. From (16), we find that $\frac{\partial{\bm\gamma_{D_i}(\bar{\bm{\rho_i}},\rho_i)}}{\partial{\rho_i}}
\mid_{{\rho_i}=0}>0$ and $\frac{\partial{\bm\gamma_{D_i}(\bar{\bm{\rho_i}},\rho_i)}}{\partial{\rho_i}}
\mid_{{\rho_i}=1}<0$. Thus, there exists a $\rho_i\in[0,1]$ that satisfies  $\frac{\partial{\bm\gamma_{D_i}
(\bar{\bm{\rho_i}},\rho_i)}}{\partial{\rho_i}}=0$.
We note that the closed-form solution of $\rho_i$ to $\frac{\partial{\bm\gamma_{D_i}
(\bar{\bm{\rho_i}},\rho_i)}}{\partial{\rho_i}}=0$ cannot be obtained. Instead, we propose an iterative algorithm to find the optimal $\bm{\rho}$. The main idea of our iterative algorithm is briefly summarized as follows.  Assume that the result of $n^{\text{th}}$ iteration is ${\bm{\rho}}(n)=[{\rho_1}(n), {\rho_2}(n),\cdots,{\rho_K}(n)]$. The solution of $(n+1)^{\text{th}}$ PS ratio $\rho_i(n+1)$ is then
acquired by using Newton iteration method with the knowledge of $\{\rho_1(n+1),\rho_2(n+1),\cdots,\rho_{i-1}(n+1),\rho_{i+1}(n)
,\cdots,\rho_K(n)\}$. We define a parameter $\varepsilon$ to evaluate whether the algorithm is convergent or not. The iteration converges if and only if the condition of ${\|{\bm\rho(n+1)}-{\bm\rho(n)}\|}_2\ll{\varepsilon}$ is satisfied. 

In what follows, we determine the wise iterative initial values to reduce the number of the iterations to find the optimal $\bm{\rho}$. To this end, we obtain the optimal PS ratios under the asymptotic scenario of high SNR, which in fact serve as the wise initial values of our iterative algorithm under general scenarios.
We note that (12) can be simplified when the received SINRs at relays are high, i.e.,
$\rho_ip_i{b_{ii}}\gg{\sigma}^2$ and $\rho_jp_j{b_{jj}}\gg{\sigma}^2$. The second item and the third item of the
denominator of $\bm\gamma_{D_i}(\bar{\bm{\rho_i}},\rho_i)$ approximate to zero under the high SINR consideration.
Then, $\bm\gamma_{D_i}(\bar{\bm{\rho_i}},\rho_i)$ can be rewritten as
\begin{align}
&\tilde{\bm\gamma}_{D_i}(\bar{\bm{\rho_i}},\rho_i) \nonumber\\
=&\frac{{\eta}\left(1-\rho_i\right)X_i{\rho_i}{p_i}Y_{ii}
{\frac{1}{\rho_ip_i{b_{ii}}+{\sigma}^2}}}{\sum\limits_{j=1,j\neq{i}}^K{\eta}\left(1-\rho_j\right)X_j{\rho_j}{p_j}Y_{ji}
{\frac{1}{\rho_ip_i{b_{ii}}+{\sigma}^2}}+{\sigma^2}}.
\setcounter{equation}{18}
\end{align}
The first-order derivative of $\tilde{\bm\gamma}_{D_i}(\bar{\bm{\rho_i}},\rho_i)$ with respect to $\rho_i$ is given by
\begin{align}
&\frac{{\tilde{\bm\gamma}_{D_i}(\bar{\bm{\rho_i}},\rho_i)}}{\partial{\rho_i}} \nonumber\\
&=\frac{X_iY_{ii}{p_i}\left(s_{ji}+\sigma^2\right)\left(-{\rho_i}^2p_i{b_{ii}}-2\rho_i{\sigma}^2+{\sigma}^2\right)}
{\left[\rho_i{p_i}{b_{ii}}\left(s_{ji}+{\sigma}^2\right)+{\sigma}^2s_{ji}+{\sigma}^2\right]^2},
\end{align}
where $s_{ji}={\sum\limits_{j=1,j\neq{i}}^K}{{\eta}\left(1-\rho_j\right)}{X_i}{\rho_j}{p_j}{Y_{ji}}{\frac{1}{\rho_jp_i{b_{jj}}+\sigma^2}}$.
We find that $\frac{{\tilde{\bm\gamma}_{D_i}(\bar{\bm{\rho_i}},\rho_i)}}{\partial{\rho_i}}\mid_{\rho_i=0}>0$ and  $\frac{{\tilde{\bm\gamma}_{D_i}(\bar{\bm{\rho_i}},\rho_i)}}{\partial{\rho_i}}\mid_{\rho_i=1}<0$.
Thus, there exists a ${\rho_i}\in{[0,1]}$ satisfying that $\tilde{\bm\gamma}_{D_i}(\bar{\bm{\rho_i}},\rho_i)$ in (20) is equal to zero. In addition, we need to solve the following equation to acquire the initial PS ratios.
\begin{align}
\hat{f}(\rho_i)=X_iY_{ii}p_i\left(s_{ji}+{\rho_i}^2\right)\left(-{\rho_i}^2{p_i}{b_{ii}}-2{\rho_i}{\rho_i}^2+{\rho_i}^2\right)=0.
\end{align}
Since $b_{ii}={r_{ii}^{-\tau}}||{\tilde{\mathbf{H}}_{ii}}||^2$, it is always true that $p_i{b_{ii}}>0$.
Note that the solution of (21) is uncorrelated with $\bar{\bm{\rho_i}}$. Hence, each  $\rho_i$ can be calculated independently. We further find that (21) has two real roots, since $\Delta=4\sigma^2
(\sigma^2+p_ib_i)>0$. Moreover, according to the curve of the function as  depicted in Figure~3, the solution of $\rho_i=\frac{-2\sigma^2-\sqrt{\Delta}}{2p_i{b_{ii}}}<0$ should be abandoned, and the optimal PS ratio under the high SINR condition is given by $\rho_i^o=\frac{-2\sigma^2+\sqrt{\Delta}}{2p_i{b_{ii}}}$.

\begin{figure}[!htb]
\centering
\includegraphics[width=0.85\columnwidth]{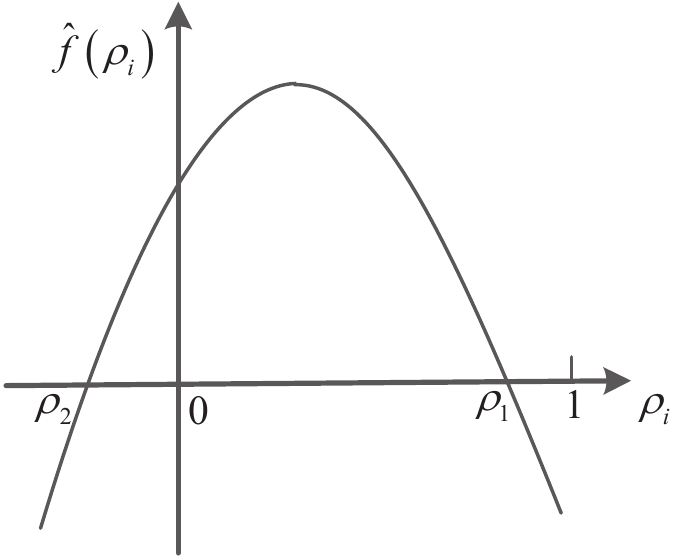}
\caption{The curve of $\hat{f}(\rho_i)$ versus $\rho_i$.}
\label{Fig:function1}\vspace{9mm}
\end{figure}

Therefore, the initial PS ratio for the $i^{\text{th}}$ link of our iterative algorithm is given by
\begin{equation}
{\rho_i(0)}=\rho_i^o=\frac{-2{\sigma^2}+\sqrt{\Delta}}{2p_i{b_{ii}}}.
\end{equation}

Different from the system of linear equations, it is  difficult to provide a convergence condition for our proposed iterative algorithm. However, it can be found that ${{\rho_i}(n+1)}$ is the only root of equation ${\bm\gamma}_{D_i}(\bar{\bm{\rho_i}},\rho_i)$. Thus, the result of every iteration increases the objective function, i.e., ${\bm\gamma}_{D_i}({\rho_i}(n+1))>{\bm\gamma}_{D_i}({\rho_i}(n))$. Considering the monotonic increase of the objective function and the fact that the maximum achievable $\bm{\gamma}_{D_i}$ is finite, the proposed algorithm converges to a certain point, at which the iteration stops.
The detailed steps of the proposed algorithm are given in Table~\ref{table:1}.

\begin{table}[!t]
\centering
\caption{Algorithm to solve (P2)}\label{table:1}
\normalsize
\begin{tabular}{lcl}
\smallskip
1. \textbf{Initialization:} Set iteration number $n=0$; Determine the
\\ \smallskip
initial PS ratios $\bm\rho(0)=\{{\rho_1(0)},{\rho_2(0)}\cdots{\rho_K(0)}\}$ by (22). \\
\smallskip
2. \textbf{Iteration:} Obtain the $n^{\text{th}}$ iteration results   ${\bm\rho}(n+1)$ by \\ \smallskip
solving (16) for each $\rho_i$ with the Newton iteration method \\ \smallskip
and $\{{\rho_1(n+1)}\cdots{\rho_{i-1}(n+1)},{\rho_{i+1}(n)}\cdots{\rho_K(n)}\}$.\\
\smallskip
3. Check if ${\|{\bm\rho(n+1)}-{\bm\rho(n)}\|}_2\ll{\varepsilon }$ is satisfied. If it is \\ \smallskip
 satisfied, go to Step 4. Otherwise, update the iteration \\ \smallskip
 number $ n=n+1$  and return to Step 2. \\ \smallskip
4. Obtain the optimal PS ratios as
${\bm\rho}^*=\bm\rho(n+1)$. \\
\end{tabular}
\end{table}

\vspace{2mm}
\section{Numerical Results}
In this section, we show the performance of the proposed scheme by simulations. 
Unless otherwise stated, we set the distance between each source and its dedicated relay $r_{ii}=1$, the distance between each relay and its respective destination $ m_{ii} = 1$,  
the path-loss exponent $\tau=3$,
the energy conversion efficiency $\eta=0.5$,
the variance of AWGN at receivers $\sigma^2=0.01$,
and the convergency criteria  $\varepsilon = 10^{-3}$.
We randomly generate 10,000 channel realizations to obtain the average sum rate of the network.
\begin{figure}[!t]
\centering
\includegraphics[width=3.45in]{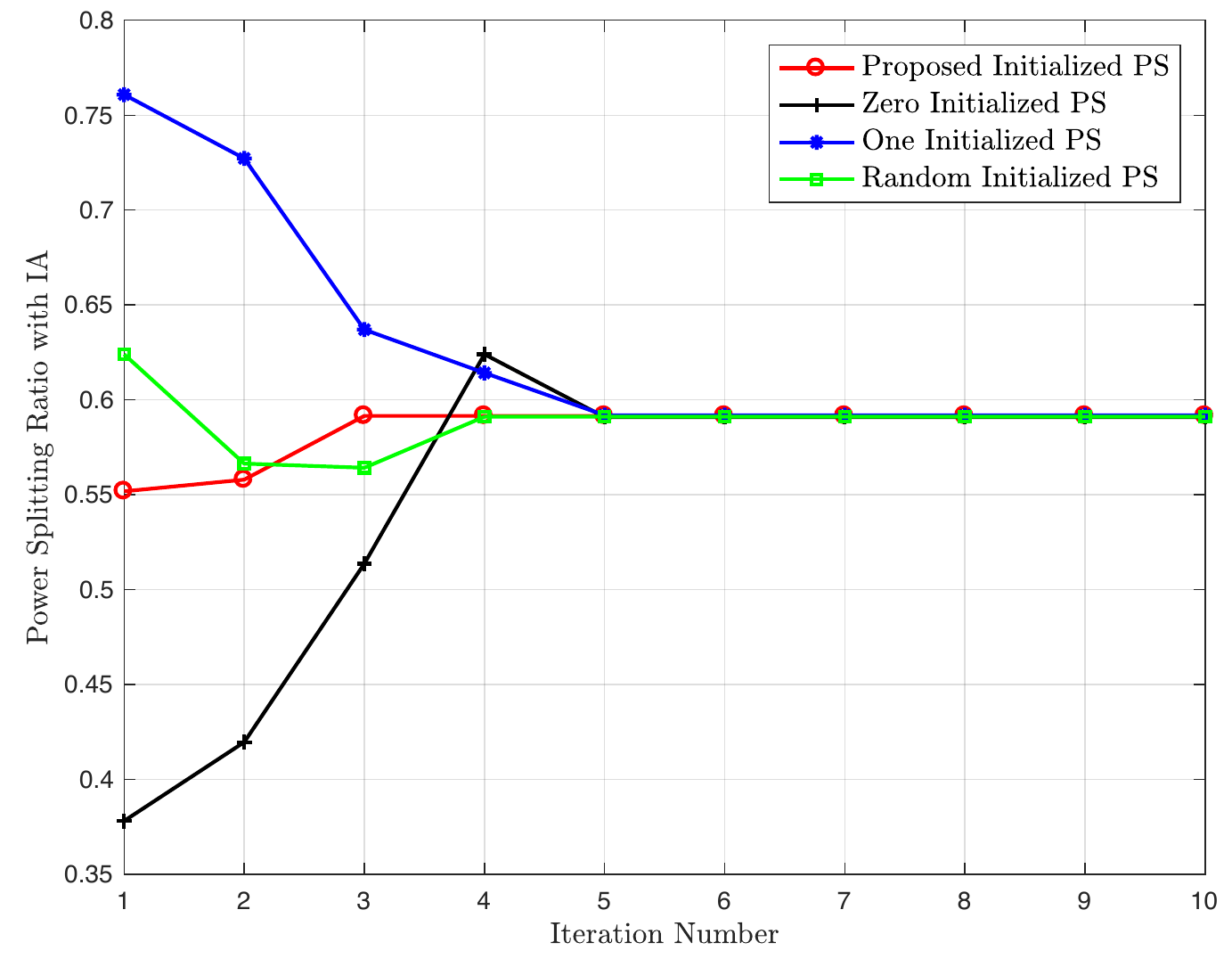}
\caption{Illustration of the convergence of the proposed algorithm. The system parameters are $M=N=L=4$ and $K=3$.}
\end{figure}
\vspace{2mm}

We first demonstrate the convergence speed of the proposed iterative algorithm.
Figure~4 plots the obtained PS ratio versus the iteration number for
different initial values.
As shown in the figure, the algorithm always converges to the same value of the PS ratio, i.e., the optimal PS ratio, for any given initial values. Comparing the results for different initial points, we find that our proposed initialized PS scheme outperforms all other schemes, in terms of the use of iterations to obtain the optimal PS ratio.

We now present the performance comparison between our proposed scheme and benchmark schemes. We consider three benchmark schemes, which are the transmission with randomly select PS ratios, the transmission with fixed PS ratios of $\rho_i=0.5$, and the transmission scheme without the IA as proposed in~\cite{7ChenH}.
Figures~5 and 6 plot the average sum rate versus the transmit power at the source and the number of links, respectively.
As shown in both figures, our proposed scheme always significantly outperforms all benchmark schemes. We note that the benchmark scheme without the IA always has the worst performance, which implies that the introduction of the IA improves the network performance even if the PS ratios are not optimally designed.
The great advantage of our proposed scheme is due to not only the introduction of the IA to eliminate the interference to relays but also the well-designed PS ratios that balance the information processing and energy harvesting to forward the information.

%

We next discuss the impact of the transmit power at the source on the performance of the network.  
As depicted in Figure~5,  the average sum rate increases as the transmit power increases.
From all curves, we note that the speed of the increase of sum rate decreases as the transmit power increases, and it becomes very slow when the transmit power is high.
Such an observation can be explained as follows.
In this work, we have not employed the IA for the second hop due to the practical consideration of the system overhead and the high energy consumption at relays. Then, the increase of transmit power at relays may harm the network performance due to the interference. Thus, the relays actually cannot always increase the transmit power for the second hop, even if the harvested energy increases.
Furthermore, we find that the performance gap between our proposed scheme and the other benchmark schemes increases as the transmit power at the source increases. This is because that the proposed scheme wisely takes advantage of the increase of the available energy resource at sources by the IA and the PS design, while the other schemes cannot take fully advantage of the increase of resource due to lack of either interference management or the PS design.

%
\begin{figure}[tb]
\centering
\includegraphics[width=3.45in]{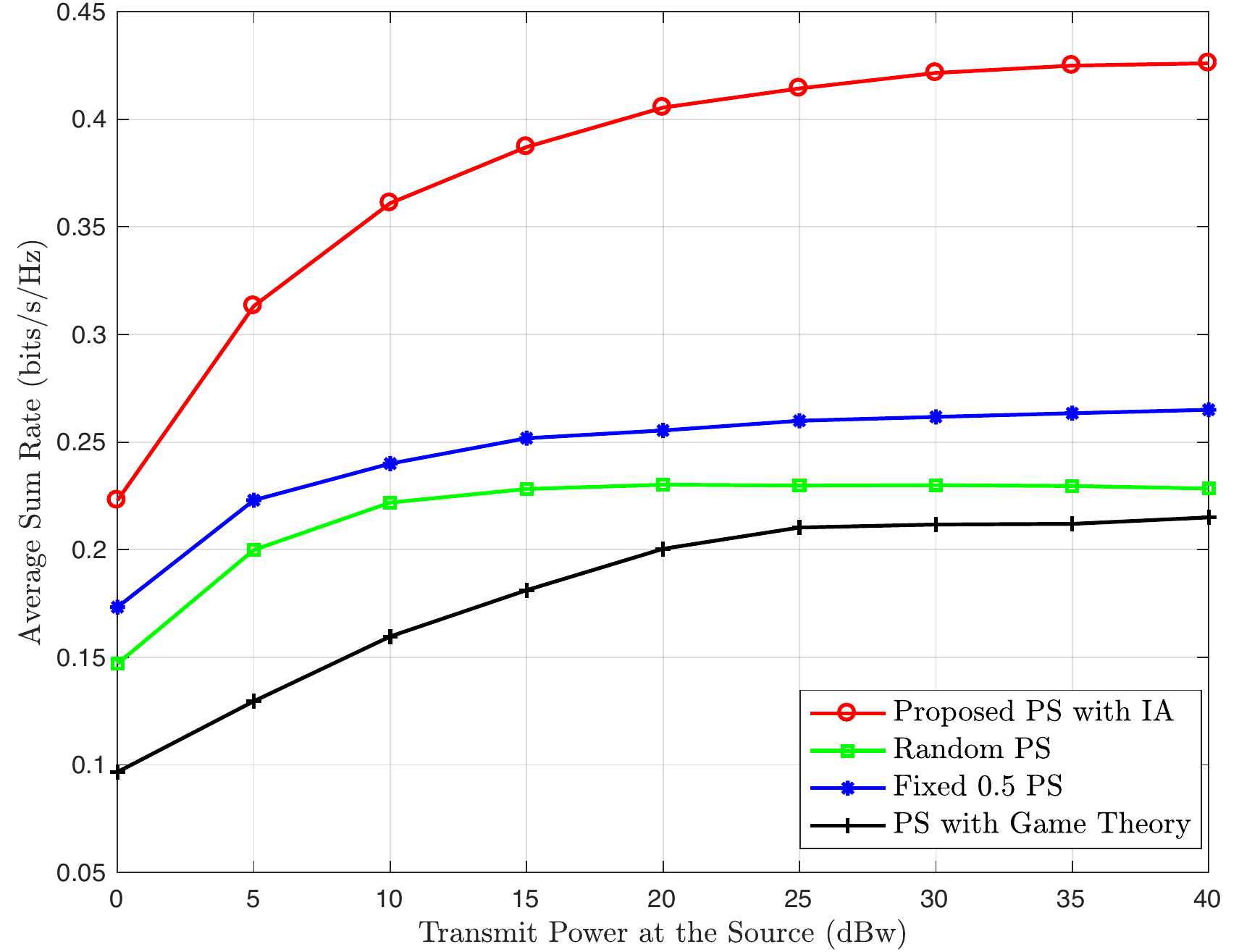}
\caption{Sum rate of the network versus transmit power at the source. The system parameters are $M=N=L=K=4$.}\vspace{4mm}
\end{figure}
\begin{figure}[t]
\centering
\includegraphics[width=3.45in]{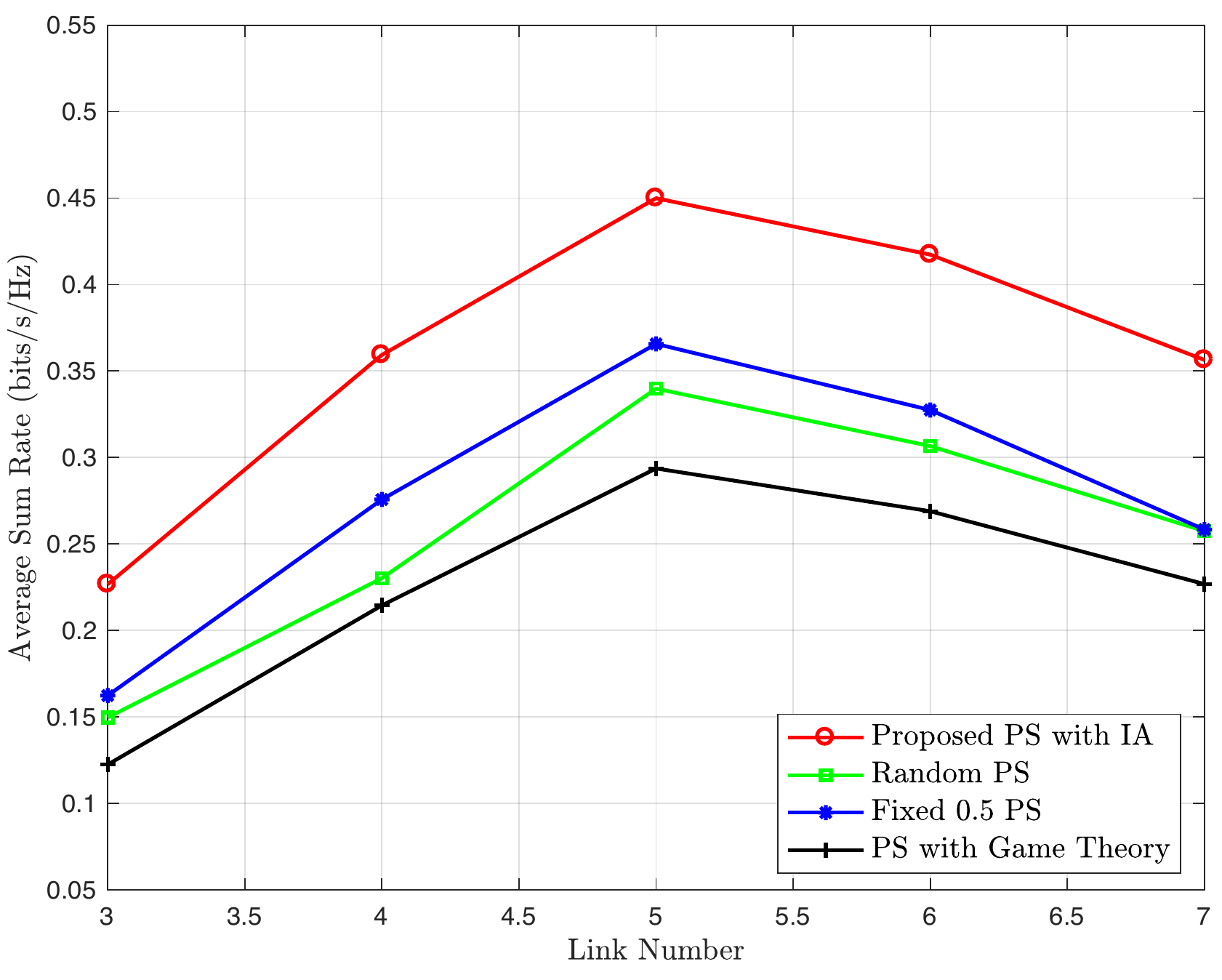}
\caption{Sum rate of the network versus number of links. The system parameters are $M=N=L=4$.}\vspace{2mm}
\end{figure}

Finally, we evaluate the impact of the number of links on  the performance of the network. As illustrated in Figure~6, as the number of links increases, the sum rate of the network increases first until achieving a peak value, and then decreases. We explain this observation as follows.
When the number of links is small, the increase of the number of links can significantly increase the multiplexing gain of the network with the relatively small interference. When the number of links becomes large, further increasing
the number of links would make the interference issue become serious, as the IA has not been employed at the relays for the second-hop transmission in the proposed scheme.
From the discussions of Figures 5 and 6, we note that adopting the IA at the second hop has a considerable potential to further improve the performance of the network. However, as mentioned previously, the employment of the IA at relays would considerably increase the system overhead and the energy consumption of the energy-constrained relays. Thus, it is not always feasible to employ the IA at the wireless-powered relays in practice.

\vspace{2mm}
\section{Conclusion}
In this paper, we have introduced a novel transmission scheme for multi-user multi-relay IC networks, where the IA is employed at the sources and the PS is designed at the AF relays. We have proposed a distributed and iterative algorithm to determine the optimal PS ratios that maximize the sum rate of the network.
In order to reduce the number of iterations to obtain the optimal PS ratios, we have further derived the closed-form expressions for the optimal PS ratios under the high SNR condition, which serve as the initial values of the iteration for the proposed algorithm.
Our results have shown that the proposed scheme significantly outperforms the benchmark schemes to improve the network performance.

\vspace{2mm}
\section*{Acknowledgment}
This work is supported by National Natural Science Foundation of China (NSFC) under grant 61461136001.

\end{document}